\documentclass[aps,prd,preprintnumbers,showpacs,showkeys,nofootinbib,
superscriptaddress,fleqn,floatfix,tightenlines]{revtex4-1}
\usepackage{amsmath,amsfonts,amssymb,amscd,amsxtra,amsthm}
\usepackage{graphicx}  
\usepackage{epstopdf}
\usepackage{dcolumn}  
\usepackage{bm}          
\usepackage{slashed}
\usepackage[utf8]{inputenc} 
\usepackage[normalem]{ulem} 
\usepackage[dvipsnames]{xcolor} 
\usepackage{tabularx}
\usepackage{enumitem}  
\usepackage{array} 
\usepackage{slashed}
\usepackage{tikz}
\usepackage{cleveref} 
\usepackage{multirow}
\renewcommand\sout{\bgroup \color{red} \ULdepth=-.5ex \ULset}

\makeatletter

\begin{document}  
\preprint{INHA-NTG-07/2018}
\title{Meson-baryon coupling constants of the SU(3) baryons with
  flavor SU(3) symmetry breaking}

\author{Ghil-Seok Yang}
\email[E-mail: ]{ghsyang@ssu.ac.kr}
\affiliation{Department of Physics, Soongsil University, Seoul 06978,
  Republic of Korea}

\author{Hyun-Chul Kim}
\email[E-mail: ]{hchkim@inha.ac.kr}
\affiliation{Department of Physics, Inha University, Incheon 22212,
  Republic of Korea}
\affiliation{School of Physics, Korea Institute for Advanced Study
  (KIAS), Seoul 02455, Republic of Korea}
\date{\today}
\begin{abstract}
We investigate the strong coupling constants for the baryon octet-octet,
decuplet-octet, and decuplet-decuplet vertices with pseudoscalar mesons
within a general framework of the chiral quark-soliton model, taking
into account the effects of flavor SU(3) symmetry breaking to linear
order in the expansion of the strange current quark mass. All relevant
dynamical parameters are fixed by using the experimental data on hyperon
semileptonic decays and the singlet axial-vector constant of the nucleon.
The results of the strong coupling constants for the baryon octet
and the pseudoscalar meson octet are compared with those determined
from the J\"ulich-Bonn potential and the Nijmegen extended soft-core
potential for hyperon-nucleon scattering. The results of the strong
decay widths of the baryon decuplet are in good agreement with the
experimental data. The effects of $SU_f(3)$ symmetry breaking are 
sizable on the $\eta'$ coupling constants. We predict also the strong
coupling constants for the $\Omega$ 
baryons.  
\end{abstract}

\maketitle


\section{Introduction}

The meson-baryon coupling constants are the essential quantities in
understanding the structure of SU(3) baryons and in describing various
productions such as meson-baryon scattering, baryon-baryon scattering,
photoproduction and electroproduction of hadrons. The strong coupling
constants are often determined with flavor $\mathrm{SU}_{f}(3)$ symmetry
assumed. Knowing the $\pi NN$ coupling constants and the ratio
$\alpha=F/(F+D)$, where $F$ and $D$ are the two couplings arising from
the SU(3) Wigner-Eckart theorem for computing the matrix elements of
the axial-vector current~\cite{deSwart:1963gc}, one can determine the
pseudoscalar meson octet ($P_{8}$) and baryon octet ($B_{8}$) coupling
constants:  
\begin{align}
g_{P_{8}^{i}B_{8}^{j}B_{8}^{k}}
\;=\;
g\left[i\alpha
  f_{ijk}+(1-\alpha)d_{ijk}\right],
\label{eq:1}
\end{align}
with $g=g_{\pi NN}$. The ratio $\alpha$ can be found from the five
known experimental data on hyperon semileptonic decay (HSD) constants
$(g_{1}/f_{1})^{B_{8}'\to B_{8}}$~\cite{PDG,Goto:1999by}. Almost
all theoretical works on the hyperon-nucleon interaction use it obtained
in this way~\cite{Holzenkamp:1989tq,Reuber:1995vc,Rijken:1998yy, 
Haidenbauer:2005zh,Rijken:2010zzb,Haidenbauer:2013oca}.
However, the empirical values of $F$ and $D$ determined from the
HSD constants contain tacitly the effects of flavor $\mathrm{SU}_{f}(3)$
symmetry breaking, though $F$ and $D$ are defined with $\mathrm{SU}_{f}(3)$
symmetry assumed.

The strong coupling constants for the baryon decuplet ($B_{10}$)-octet
and pseudoscalar meson octet vertices are less known. Even the $\pi N\Delta$
coupling constant, which is the essential quantity in describing the
$NN$ and $\pi N$ interactions, is not at all given in consensus.
The $\pi N\Delta$ coupling constant is usually determined by the
decay width of $\Delta\to\pi N$, which yields $f_{\pi N\Delta}\approx2.24$.
In describing $\pi N$ scattering, $f_{\pi N\Delta}\approx2.0-2.5$
was used~\cite{Schutz:1994wp,Sato:1996gk,Polinder:2005sn}. On the
other hand, the full Bonn potential for the $NN$
interaction~\cite{Machleidt:1987hj}, $f_{\pi N\Delta}=1.678$ was
employed, which was taken from the relation in an SU(6) quark model
$f_{\pi N\Delta}^{2}=72f_{\pi NN}^{2}/25$~\cite{Brown:1975di}. A
recent work determined $f_{\pi N\Delta}=1.256$, which is much smaller
that that from the decay width, based on the global fit to the $\pi N$
and $\gamma N$ data~\cite{Kamano:2013iva}. When it comes to the
coupling constants for the other members of the baryon decuplet,
information is much less known. 

In the mean time, new experimental programs with strangeness of $S=-3$
are now under way at the J-PARC~\cite{Takahashi:2013eya} and a new
excited $\Omega$ resonance was reported by the Belle
Collaboration~\cite{Yelton:2018mag}. The HAL Collaboration in lattice
QCD predicted the dibaryon ($\Omega\Omega$) with strangeness
$S=-6$~\cite{Gongyo:2017fjb}. The $N\Omega$ interaction was studied in
a meson-exchange picture~\cite{Sekihara:2018tsb} very recently. The
baryon decuplet and octet interactions were
investigated~\cite{Haidenbauer:2017sws}. In this regard, it is highly 
required to provide information on the baryon and pseudoscalar meson
coupling constants in a quantitative manner. 

In the present work, we want to study the coupling constants for the
vertices of the baryon decuplet-octet (also decuplet) and pseudoscalar
mesons in a pion mean-field approach that is often called the the
chiral quark-soliton model ($\chi$SM). In
Refs.~\cite{Yang:2010fm,Yang:2011qe}, we reexamined 
the mass splittings of the SU(3) baryon octet and the decuplet, fixing
all the parameters unequivocally to the experimental data. The effects 
of $\mathrm{SU}_{f}(3)$ symmetry breaking and isospin symmetry breaking
due to both the electromagnetic interaction and current quark mass
difference~\cite{Yang:2010fm,Yang:2010id} were systematically
included, which made it possible to exploit the experimental data to
fix the parameters. Since we have fixed all unknown parameters in the
baryon wavefunctions, we can proceed to the study of the axial-vector
transitions, again fixing relevant parameters by utilizing the
experimental data on the HSD constants and the flavor-singlet
axial-vector charge $g_{A}^{(0)}$. Though similar works were done  
already~\cite{Kim:1999uf,Kim:2000by,Kim:2000vn,Yang:2007yj}, 
it was then not possible to fix all the parameters unambiguously 
because of the absence of isospin symmetry breaking which is
inevitable in incorporating the experimental data for the baryon
octet. Recently, we have shown that all the relevant parameters for
the HSD constants can be fixed without any
ambiguity~\cite{Yang:2015era}. Once they are known, we can compute all
possible axial-vector transitions between the baryon multiplets. As a
result, we are able to determine the coupling constants for the
vertices of the baryon decuplet-octet (decuplet) and pseudoscalar
mesons without any additional parameters introduced, taking into
account the effects of explicit SU(3) symmetry breaking.

This paper is outlined as follows: Section 2, we briefly review the
general formalism of the $\chi$SM to compute the axial-vector transitions
between the baryon multiplets and show how to fix the parameters for
the axial-vector transitions. In Sec. 3, we present the results 
of the coupling constants for the baryon multiplets and pseudoscalar
meson vertices. We show also the decay widths of the baryon decuplet
to the octet. In Sect. 4, we discuss the results for the $\eta$
($\eta'$), and baryon coupling constants, applying a usual mixing
between the octet $\eta_8$ and the singlet $\eta_0$. In the final
Section we summarize the present work and draw conclusions.  


\section{Baryon matrix elements of the 
axial-vector currents \label{sec:2}}

The baryon matrix elements of the axial-vector currents
are expressed in terms of three form factors 
\begin{align}
\langle B_{8}|A_{\mu}^{i}|B_{8}'\rangle 
 &=
\bar{u}_{B_{8}'}(p_{2},s_{2})\left[g_{1}^{B_{8}'
\to B_{8}}(q^{2})\gamma_{\mu}
+\frac{ig_{2}^{B_{8}'\to
    B_{8}}(q^{2})\sigma_{\mu\nu}q^{\nu}}{M_{B_{8}'}}
+\frac{g_{3}^{B_{8}'\to B_{8}}(q^{2})q_{\mu}}{M_{B_{8}'}}\right]
\gamma_{5}u_{B_{8}}(p_{1},s_{1}),
\label{eq:2}
\end{align}
where the axial-vector currents are defined as 
\begin{align}
A_{\mu}^{i}(x)
\;=\; \bar{\psi}(x)\gamma_{\mu}\gamma_{5}\frac{1}{2}\lambda^{i}\psi(x).
\label{eq:VAcurrent}
\end{align}
The $\lambda^{i}$ stand for flavor Gell-Mann matrices for strangeness
conserving $\Delta S=0$ transitions
($i\,=\,3,\,8,\,\left(1\,\pm\,i\,2\right)$) and for $|\Delta S|=1$
ones ($i\,=\,4\,\pm\,i\,5$), respectively. The $q^{2}=-Q^{2}$ denotes
the square of the momentum transfer $q=p_{2}-p_{1}$. The form factors
$g_{i}$ are real quantities due to $CP$-invariance, depending only on
the square of the momentum transfer. We can neglect 
$g_{3}^{B_{8}'\to B_{8}}$, because its contribution to the decay
rate is proportional to the ratio $m_{l}^{2}/M_{B_{8}'}^{2}\ll1$,
where $m_{l}$ represents a mass of the lepton ($e$ or $\mu$) in the
final state and that of the baryon in the initial state, $M_{B_{8}'}$,
respectively. The $g_{2}^{B_{8}'\to B_{8}}$ is finite only with the
effects of $\mathrm{SU}_{f}(3)$ symmetry and isospin symmetry breakings
because of its opposite $G$ parity to the axial-vector current, so
it is very small for the baryon octet. 

In the $\chi$QSM, the collective operator for the axial-vector constants
can be defined in terms of the $\mathrm{SU}_f(3)$ Wigner $D$ functions
~\cite{Kim:1999uf,Kim:2000by,Kim:2000vn}: 
\begin{align}
\hat{g}_{1} 
& = a_{1}\,D_{X3}^{(8)}
\;+\;a_{2}d_{pq3}\,D_{Xp}^{(8)}\,\hat{J}_{q}
\;+\;\frac{a_{3}}{\sqrt{3}}\,D_{X8}^{(8)}\,\hat{J}_{3}
+\frac{a_{4}}{\sqrt{3}}\,d_{pq3}D_{Xp}^{(8)}\,D_{8q}^{(8)}
\cr
& +
   a_{5}\,\left(D_{X3}^{(8)}\,D_{88}^{(8)}+D_{X8}^{(8)}\,D_{83}^{(8)}\right)
+ a_{6}\,\left(D_{X3}^{(8)}\,D_{88}^{(8)}-D_{X8}^{(8)}\,D_{83}^{(8)}\right),
\label{eq:4}
\end{align}
where $a_{i}$ denote dynamical parameters encoding the specific
dynamics of a
$\chi$QSM~\cite{Christov:1995vm,Kim:1997ts,Ledwig:2008ku}. 
Note that $a_{1}$ parametrizes the leading-order contribution,
$a_{2}$ and $a_{3}$ come from the rotational $1/N_{c}$ corrections,
and $a_{4}$, $a_{5}$ and $a_{6}$ are originated from $\mathrm{SU}_{f}(3)$
symmetry breaking, in which the strange current quark mass $m_{\mathrm{s}}$
is contained. $\hat{J}_{q}$ and $\hat{J}_{3}$ stand for the $q$-th and 
third components of the collective spin operator of the baryons,
respectively. The $D_{ab}^{(8)}$ are the SU(3) Wigner $D$ functions
in the octet representation.

The baryon wavefunctions for the baryon octet and decuplet are written
in terms of the $\mathrm{SU}_f(3)$ Wigner $D$ functions in the 
$\chi$SM~\cite{Yang:2010fm,Blotz:1992pw}: 
\begin{align}
\langle A|\mathcal{R},\,B(Y\,T\,T_{3},\;Y^{\prime}\,J\,J_{3})\rangle
&=\Psi_{(\mathcal{R}^{*}\,;\,Y^{\prime}\,J\,J_{3})}^{(\mathcal{R\,};\,Y\,T\,T_{3})}(A)
  \cr 
&=\sqrt{\textrm{dim}(\mathcal{R})}\,(-)^{J_{3}+Y^{\prime}/2}\,
D_{(Y,\,T,\,T_{3})(-Y^{\prime},\,J,\,-J_{3})}^{(\mathcal{R})*}(A),
\label{eq:5}
\end{align}
where $\mathcal{R}$ designates the allowed irreducible representations
of the $\mathrm{SU}_{f}(3)$ group, i.e. $\mathcal{R}\,=\,8,\,10,\,\cdots$.
$Y,\,T,\,T_{3}$ denote the corresponding hypercharge, isospin and
its third component, respectively. The right hypercharge is constrained
to be $Y^{\prime}=1$ in such a way that it selects a tower of allowed
$\mathrm{SU}_{f}(3)$ representations. The baryon octet and decuplet,
which are the lowest representations, coincide with those of the quark
model. This has been considered as a success of the collective
quantization and gives a hint about certain duality between the chiral
soliton picture and the constituent quark model.

When the effects of $\mathrm{SU}_{f}(3)$ symmetry breaking are taken
into account, a baryon state is no more pure state but the state mixed
with those in higher representations. Thus, the wavefunctions for
the baryon octet and the decuplet are given by 
\begin{align}
\left|B_{8}\right\rangle  
& =
\left|8_{1/2},B\right\rangle 
\;+\;
c_{\overline{10}}^{B}\left|\overline{10}_{1/2},B\right\rangle 
\;+\;
c_{27}^{B}\left|27_{1/2},B\right\rangle ,
\cr
\left|B_{10}\right\rangle  
& =
\left|10_{3/2},B\right\rangle 
\;+\;
a_{27}^{B}\left|27_{3/2},B\right\rangle 
\;+\;
a_{35}^{B}\left|35_{3/2},B\right\rangle ,
\label{eq:6}
\end{align}
where the spin indices $J_{3}$ have been dropped from the states.
The mixing coefficients in Eq.(\ref{eq:6}) contain the strange current
quark mass $m_{\mathrm{s}}$ and are expressed as 
\begin{align}
c_{\overline{10}}^{B}=c_{\overline{10}}\left[\kern-0.5em 
\begin{array}{c}
\sqrt{5}\\
0\\
\sqrt{5}\\
0
\end{array}\kern-0.2em \right]\kern-0.2em ,
\;c_{27}^{B}=c_{27}\left[\kern-0.5em 
\begin{array}{c}
\sqrt{6}\\
3\\
2\\
\sqrt{6}
\end{array}\kern-0.2em \right]\kern-0.2em ,
\;a_{27}^{B}=a_{27}\left[\kern-0.5em 
\begin{array}{c}
\sqrt{15/2}\\
2\\
\sqrt{3/2}\\
0
\end{array}\kern-0.2em \right]\kern-0.2em ,
\;a_{35}^{B}=a_{35}\left[\kern-0.5em 
\begin{array}{c}
5/\sqrt{14}\\
2\sqrt{5/7}\\
3\sqrt{5/14}\\
2\sqrt{5/7}
\end{array}\kern-0.2em \right],
\label{eq:7}
\end{align}
respectively in the bases $[N,\;\Lambda,\;\Sigma,\;\Xi]$ and
$[\Delta,\;\Sigma^{\ast},\;\Xi^{\ast},\;\Omega]$ with 
\begin{align}
c_{\overline{10}} & =
-\frac{I_{2}}{15}\left(m_{s}-\hat{m}\right)
\left(\alpha+\frac{1}{2}\gamma\right),
\;\;\;\;\;
c_{27}=-\frac{I_{2}}{25}\left(m_{s}-\hat{m}\right)
\left(\alpha-\frac{1}{6}\gamma\right),
\cr
a_{27} &
=-\frac{I_{2}}{8}\left(m_{s}-\hat{m}\right)
\left(\alpha+\frac{5}{6}\gamma\right),
\;\;\;\;\;
a_{35}=-\frac{I_{2}}{24}\left(m_{s}-\hat{m}\right)
\left(\alpha-\frac{1}{2}\gamma\right),
\cr
d_{8} &
=\frac{I_{2}}{15}\left(m_{s}-\hat{m}\right)
\left(\alpha+\frac{1}{2}\gamma\right),
\;\;\;\;\;\;\;
d_{27}\;=\;-\frac{I_{2}}{8}\left(m_{s}-\hat{m}\right)
\left(\alpha-\frac{7}{6}\gamma\right),
\cr
d_{\overline{35}} & =
-\frac{I_{2}}{4}\left(m_{s}-\hat{m}\right)
\left(\alpha+\frac{1}{6}\gamma\right).
\label{eq:8}
\end{align}
Here $I_{2}$ is a moment of inertia for the soliton. $\alpha$ and
$\gamma$ are the parameters appearing in the collective Hamiltonian.
As for the explicit definitions of $I_{2}$, $\alpha$ and $\gamma$,
we refer to Ref.~\cite{Yang:2010fm}, where one can find also a detailed
discussion as to how they are fixed unambiguously, and relevant references.

Since the baryon wavefunctions contain the corrections of linear
$\mathrm{SU}_{f}(3)$ symmetry breaking as shown in Eq.~(\ref{eq:6}),
the axial-vector transition constants $g_{1}^{B_{8}'\to B_{8}}$
acquire yet another linear $m_{\mathrm{s}}$ corrections, when the
collective operator $\hat{g}_{1}$ is sandwiched between the baryon
states. Thus, we have the two different linear $m_{\mathrm{s}}$
corrections to the axial-vector transition constants, i.e., one from
$a_{4}$, $a_{5}$ and $a_{6}$, and the other from the baryon
wavefunctions. 
\begin{table}[htp]
\caption{The experimental data on the hyperon semileptonic decay
  constants~\cite{PDG} and the singlet axial-vector constant taken
  from Ref.~\cite{Bass:2004xa}.}
\begin{tabular}{llc}
\hline 
& Experimental data  
& References
\cr
\hline 
$g_{1}/f_{1}\left(n\to p\right)$  
& $1.2723\pm0.0023$  
& PDG~\cite{PDG}
\cr
$g_{1}/f_{1}\left(\Lambda\to p\right)$  
& $0.718\pm0.015$  
& PDG~\cite{PDG} 
\cr
$g_{1}/f_{1}\left(\Sigma^{-}\to n\right)$  
& $-0.340\pm0.017$  
& PDG~\cite{PDG}
\cr
$g_{1}/f_{1}\left(\Xi^{-}\to\Lambda\right)$  
& $0.25\pm0.05$  
& PDG~\cite{PDG} 
\cr
$g_{1}/f_{1}\left(\Xi^{0}\rightarrow\Sigma^{+}\right)$  
& $1.22\pm0.05$  
& PDG~\cite{PDG} 
\cr
$g_{A}^{(0)}$  
& $0.36\pm0.03$  
& Bass et al.~\cite{Bass:2004xa} 
\cr
\hline 
\end{tabular}
\label{tab:1} 
\end{table}
Recently, we have shown how the parameters $a_{i}$ are unequivocally
fixed in detail~\cite{Yang:2015era}. The experimental data on the
HSD constants $(g_{1}/f_{1})^{B_{8}'\to B_{8}}$ and the flavor-singlet
axial-vector charge $g_{A}^{(0)}$, listed in Table~\ref{tab:1},
will be the input for fixing $a_{i}$. The parameters $a_{i}$ are
related to the experimentally known axial-vector HSD constants and
$g_{A}^{(0)}$ in a form of the matrix equation: 
\begin{align}
\bm{g}=\bm{B}\cdot\bm{a},\label{eq:MatrixForAi}
\end{align}
where 
\begin{align}
\bm{g} & =\left(\left(g_{1}/f_{1}\right)^{n\rightarrow p},\;
\left(g_{1}/f_{1}\right)^{\Lambda\rightarrow p},\;
\left(g_{1}/f_{1}\right)^{\Sigma^{-}\rightarrow n},\;
\left(g_{1}/f_{1}\right)^{\Xi^{-}\rightarrow\Lambda},
\left(g_{1}/f_{1}\right)^{\Xi^{0}\rightarrow\Sigma^{+}},\;
g_{A}^{0}\right),
\label{eq:gmatrix}
\end{align}
\begin{align}
{\bm{B}} & =\left[\begin{array}{cccccc}
-\frac{7}{30}-\frac{1}{3}c_{\overline{10}}-\frac{2}{45}c_{27} 
& \frac{7}{60}-\frac{1}{3}c_{\overline{10}}-\frac{4}{45}c_{27} 
& \frac{1}{60}-\frac{1}{6}c_{\overline{10}}+\frac{1}{15}c_{27} 
& -\frac{11}{270} 
& -\frac{1}{9} 
& -\frac{1}{15}\\[0.3em]
-\frac{2}{15}+\frac{1}{6}c_{\overline{10}}+\frac{1}{30}c_{27} 
& \frac{1}{15}+\frac{1}{6}c_{\overline{10}}+\frac{1}{15}c_{27} 
& \frac{1}{30}+\frac{1}{12}c_{\overline{10}}-\frac{1}{20}c_{27} 
& \frac{1}{45} 
& 0 
& -\frac{1}{30}\\[0.3em]
\frac{1}{15}-\frac{1}{45}c_{27} 
& -\frac{1}{30}-\frac{2}{45}c_{27} 
& \frac{1}{15}+\frac{1}{30}c_{27} 
& -\frac{1}{270} 
& -\frac{1}{45} 
& \frac{1}{30}\\[0.3em]
-\frac{1}{30}-\frac{1}{30}c_{27} 
& \frac{1}{60}-\frac{1}{15}c_{27} 
& \frac{1}{20}+\frac{1}{20}c_{27} 
& -\frac{1}{180} 
& \frac{1}{30} 
& -\frac{1}{30}\\[0.3em]
-\frac{7}{30}+\frac{1}{6}c_{\overline{10}}+\frac{1}{45}c_{27} 
&
\frac{7}{60}+\frac{1}{6}c_{\overline{10}}+\frac{2}{45}c_{27} 
&
\frac{1}{60}+\frac{1}{12}c_{\overline{10}}-\frac{1}{30}c_{27} 
& \frac{11}{540} 
& \frac{1}{18} 
& \frac{1}{30}\\[0.3em]
0 & 0 & 1 & 0 
& -\frac{1}{5} & \frac{1}{5}\\[0.3em]
\end{array}\right],
\label{eq:BMatrix}\\
{\bm{a}} &
=\left(a_{1},\;a_{2},\;a_{3},\;a_{4},\;a_{5},\;a_{6}\right).
\label{eq:gfa}
\end{align}
Inverting $\bm{B}$, we can easily derive the parameters $a_{i}$
of which the numerical values are listed in Table~\ref{tab:2}. 
\begin{table}[htp]
\caption{Numerical values of the dynamical parameters $a_{i}$}
\begin{tabular}{cccccc}
\hline 
$a_{1}$  
& $a_{2}$  
& $a_{3}$  
& $a_{4}$  
& $a_{5}$  
& $a_{6}$ \cr
\hline 
$-3.509\pm0.011$  
& $3.437\pm0.028$  
& $0.604\pm0.030$  
& $-1.213\pm0.068$  
& $0.479\pm0.025$  
& $-0.735\pm0.040$ 
\cr
\hline 
\end{tabular}
\label{tab:2} 
\end{table}
All other unmeasured HSD constants for the baryon octet and decuplet
were predicted in Ref.~\cite{Yang:2015era}.


\section{Coupling constants for the $P_{8}$-$B_{8}$-$B_{10}$ and
  $P_{8}$-$B_{10}$-$B_{10}$ vertices~
\label{sec:10-8-8}}  

The matrix elements of the $B_{10}\rightarrow B_{8}$ and
$B_{10}\rightarrow B_{10}$ transitions with the axial-vector current
are parametrized in terms of the Adler form factors
$C_{i}^{A,B_{10}\to B_{8}}$~\cite{Adler:1968tw,
  LlewellynSmith:1971uhs, Alexandrou:2013opa} 
\begin{align}
& \left\langle
  B_{8}(p',s')\left|A_{\mu}^{i}\right|B_{10}(p,s)\right\rangle   
\cr
& 
=\;\overline{u}\,(p^{\prime},s')
\left[\left\{ \frac{C_{3}^{A,B_{10}\to
  B_{8}}(q^{2})}{M_{8}}\gamma^{\alpha}
+\frac{C_{4}^{A,B_{10}\to B_{8}}(q^{2})}{M_{8}^{2}}p^{\alpha}\right\} 
\left(q_{\alpha}g_{\mu\nu}-q_{\nu}g_{\alpha\mu}\right)\right.
\cr
 & 
\left.
+\;C_{5}^{A,B_{10}\to B_{8}}(q^{2})g_{\mu\nu}
+\frac{C_{6}^{A,B_{10}\to
  B_{8}}(q^{2})}{M_{8}^{2}}q_{\mu}q_{\nu}\right]
u^{\nu}(p,s),
\label{eq:10-8matrix}
\end{align}
\begin{align}
&\left\langle
  B_{10}(p',s')\left|A_{\mu}^{i}\right|B_{10}(p,s)\right\rangle  
\cr
& =\;\overline{u}^{\alpha}\,(p^{\prime},s')
\left[g_{\alpha\beta}\left(h_{1}(q^{2})\gamma_{\mu}\gamma_{5}
+h_{3}(q^{2})\frac{q_{\mu}}{2M_{10}}\gamma_{5}\right)\right.
\cr
 &
  \left.
+\frac{q_{\alpha}q_{\beta}}{4M_{10}^{2}}
\left(h_{1}^{\prime}(q^{2})\gamma_{\mu}\gamma_{5}
+h_{3}^{\prime}(q^{2})\frac{q_{\mu}}{2M_{10}}\gamma_{5}\right)\right]
u^{\beta}(p,s),
 \label{eq:10-10matrix}
\end{align}
where the $u^{\nu}$ represents the Rarita-Schwinger spinor for the
baryon decuplet. $q_{\mu}$ denotes the momentum transfer
$q_{\mu}=(p'-p)_{\mu}$. The axial-vector constant $C_{5}^{A,B_{10}\to
  B_{8}}$ can be related to the strong coupling constants for
$P_{8}$-$B_{8}$-$B_{10}$ and $P_{8}$-$B_{10}$-$B_{10}$ vertices by the
partially conserved axial-vector current (PCAC) hypothesis. The
pseudoscalar meson decay constant $f_8$ is defined as the transition
matrix element of the axial-vector current from the physical pion
state to the vacuum  
\begin{align}
\langle0|A_{\mu}^{a}(x)|\pi^{b}(p)\rangle=ip_{\mu}f_{8}e^{-ip\cdot
  x}\delta^{ab},
\label{eq:PionDecay}
\end{align}
which will be used for the relations of the pseudovector coupling
constants $f_{P_8B_8 B_{10}}$ and $f_{P_8B_{10} b_{10}}$ to the Adler
form factors. In the present work, we will determine only $C_5^A$ and
$h_1$.  
 
The effective Lagrangians for the $P_8 B_8 B_{10}$ and $P_8 B_{10}
B_{10}$ vertices are expressed as 
\begin{align}
\mathcal{L}_{P_{8}B_{8}B_{10}} 
& =
\frac{f_{P_{8}B_{8}B_{10}}}{m_{8}}
\overline{B}_{10}^{\mu}Z_{\mu\nu}\bm{I}\left(\frac{3}{2},\,
\frac{1}{2}\right)B_{8}\partial^{\nu}M_{8}+\mathrm{h.c.},
\cr
\mathcal{L}_{P_{8}B_{10}B_{10}} 
& =
\frac{f_{P_{8}B_{10}B_{10}}}{m_{8}}
\overline{B}_{10}^{\alpha}Z_{\alpha\beta}^{\nu}\bm{I}
\left(\frac{3}{2},\,\frac{3}{2}\right)
B_{10}^{\beta}\partial_{\nu}M_{8}+\mathrm{h.c.},
\end{align}
where the pseudovector coupling constants are defined as 
\begin{align}
f_{P_{8}B_{8}B_{10}} & =\;\frac{m_{8}}{f_{8}}C_{5}^{A}(0),
\label{eq:fm810}\\
f_{P_{8}B_{10}B_{10}} & =\;\frac{m_{8}}{f_{8}}h_{1}(0),
\label{eq:fm1010}
\end{align}
$m_{8}$ denotes the mass of the pseudoscalar meson. The field
operators $B_{10}^{\mu}$, $B_{8}$, and $P_{8}$ correspond respectively
to a decuplet baryon, a octet baryon, and a pseudoscalar octet meson.
The $Z_{\mu\nu}$ and $Z_{\alpha\beta}^{\nu}$ stand for the tensors
including the off-shell effects arising from the Rarita-Schwinger
field quantization, defined as
$Z_{\mu\nu}=g_{\mu\nu}-x_\Delta \gamma_{\mu}\gamma_{\nu}$ with 
the off-shell parameter $x_\Delta$. $\bm{I}(3/2,\,1/2)$ and
$\bm{I}(3/2,\,3/2)$ are isospin transition matrices.

For completeness, we also want to mention that the pseudoscalar strong
coupling constants can be derived from the generalized
Goldberger-Treiman (GT)
relation~\cite{General:2003sm,Alexandrou:2007xj}, which is 
defined as 
\begin{align}
g_{P_{8}B_{8}B_{10}} &
\approx\;\frac{M_{8}+M_{10}}{f_{8}}C_{5}^{A}(0).
\label{eq:axial-vector-contrib}
\end{align}
However, there is a caveat in Eq.~(\ref{eq:axial-vector-contrib}).
Keeping in mind that certain effects on the GT relation will arise
from the flavor $\mathrm{SU}_f(3)$ symmetry breaking. In
Ref.~\cite{Zhu:2002kh}, 
it was shown that loop corrections to the GT relation, which come
from the pion mass, are indeed very small ($\sim2\,\%$). So, we expect
that the strange current quark mass will not yield much effects on
the relation. Thus, as often assumed in the hyperon-nucleon potentials,
one still can use Eq.(\ref{eq:axial-vector-contrib}), if one wants
to derive the strong coupling constants $g_{P_{8}B_{8}B_{10}}$.

In effect, the numerical values of the $C_{5}^{A}(0)$ were already
presented in the previous work~\cite{Yang:2015era}. Thus, we will
show the results for the pseudovector coupling constants and decay
widths of the baryon decuplet in this work, using the experimental
data on the meson decay constants, $f_{\pi}=92.4\,\mathrm{MeV}$ and 
$f_{K}=113.0\,\mathrm{MeV}$. In Table~\ref{tab:3}, we list the results
of the pseudoscalar coupling constants for the various
$P_{8}B_{8}B_{8}$ vertices,
i.e. $g_{P_{8}B_{8}B_{8}}/\sqrt{4\pi}$. The second column represents 
those in the $\mathrm{SU}_f(3)$ symmetric case, whereas the third one
denotes those with explicit $\mathrm{SU}_f(3)$ symmetry breaking taken
into account. The results are compared with those determined from the
extended soft-core Nijmegen hyperon-nucleon ($YN$) potential
(ESC08a)~\cite{Rijken:2010zzb} and J\"ulich-Bonn $YN$ potential,
employing the generalized GT relation for kaon vertices.  
Except for the coupling constants of the vertices $\pi\Xi\Xi$ and
$K\Xi\Lambda$, the present results are in good agreement with the
those from both the Nijmegen and J\"ulich-Bonn potentials. When the
effects of the $\mathrm{SU}_f(3)$ symmetry breaking are taken into
account, the present results are more deviated from those taken from
the Nijmegen potential. Note that both the Nijmegen and J\"ulich-Bonn
potentials have assumed $\mathrm{SU}_f(3)$ symmetry and the
following relations for the $P_{8}B_{8}B_{10}$ vertices are obtained
in exact $\mathrm{SU}_f(3)$: 
\begin{align}
f_{\pi N\Delta}\;\;= & 
\sqrt{2}f_{\pi\Lambda\Sigma^{\ast}}
\;\;=\;\;-\sqrt{2}f_{\pi\Sigma\Sigma^{\ast}}
\;\;=\;\;\sqrt{2}f_{\pi\Xi\Xi^{\ast}},\cr
f_{K\Sigma\Delta}\;\;= & 
\sqrt{2}f_{KN\Sigma^{\ast}}
\;\;=\;\;-\sqrt{2}f_{K\Xi\Sigma^{\ast}}
\;\;=\;\;-\sqrt{2}f_{K\Sigma\Xi^{\ast}}\cr
= & \sqrt{2/3}f_{K\Lambda\Xi^{\ast}}
\;\;=\;\;-\sqrt{1/3}f_{K\Xi\Omega},
\label{eq:rel8810}
\end{align}
which can be found in various works already. 

\begin{table}[htp]
\caption{Pseudoscalar strong coupling constants of the baryon octet,
  divided by $\sqrt{4\pi}$. The second column lists the results for
  the   $\mathrm{SU}_f(3)$ symmetric case, whereas the third one does
  those   with explicit $\mathrm{SU}_f(3)$ symmetry breaking taken
  into   account. The fourth and fifth columns list the values of the 
  coupling constants taken from the Nijmegen and J\"ulich-Bonn
  hyperon-nucleon potentials, respectively.}
\begin{tabular}{ccccc}
\hline 
$P_{8}B_{8}B_{8}$  
& $g_{P_{8}B_{8}B_{8}}^{(0)}$  
& $g_{P_{8}B_{8}B_{8}}^{(\mathrm{total})}$  
& ESC08a \cite{Rijken:2010zzb}  
& J\"ulich-Bonn \cite{Holzenkamp:1989tq,Haidenbauer:2005zh} 
\cr
\hline 
$\pi NN$  
& $3.524\pm0.012$  
& $3.638\pm0.018$  
& $3.639$  
& $3.795$ 
\cr
$\pi\Lambda\Sigma$  
& $3.129\pm0.011$  
& $3.229\pm0.016$  
& $3.328$  
& $2.629$ 
\cr
$\pi\Sigma\Sigma$  
& $3.356\pm0.014$  
& $3.197\pm0.019$  
& $3.290$  
& $3.036$ 
\cr
$\pi\Xi\Xi$  
& $-1.240\pm0.009$  
& $-0.985\pm0.015$  
& $-1.475$  
& $\cdots$ 
\cr
$KN\Lambda$  
& $-3.185\pm0.030$  
& $-3.189\pm0.032$  
& $-3.217$  
& $-3.944$ 
\cr
$KN\Sigma$  
& $0.820\pm0.009$  
& $0.905\pm0.011$  
& $0.975$  
& $0.759$ 
\cr
$K\Lambda\Xi$  
& $1.076\pm0.013$  
& $1.316\pm0.017$  
& $0.942$  
& $\cdots$ 
\cr
$K\Sigma\Xi$  
& $-3.855\pm0.037$  
& $-3.793\pm0.037$  
& $-3.980$  
& $\cdots$ 
\cr
\hline 
\end{tabular}
\label{tab:3} 
\end{table}
\begin{table}[htp]
\caption{Pseudovector coupling constants for the $P_{8}B_{8}B_{10}$
  vertices. The second column lists the results for the
  $\mathrm{SU}_f(3)$ symmetric case, whereas the third one does those
  with explicit $\mathrm{SU}_f(3)$ symmetry breaking taken into
  account. The last column lists the values of the 
  coupling constants taken from the J\"ulich-Bonn hyperon-nucleon
  potential.} 
\begin{tabular}{cccc}
\hline 
$M_{8}B_{8}B_{10}$  
& $f_{P_{8}B_{8}B_{10}}^{(0)}$  
& $f_{P_{8}B_{8}B_{10}}^{(\mathrm{total})}$  
& J\"ulich-Bonn \cite{Holzenkamp:1989tq} 
\cr
\hline 
$\pi N\Delta$  
& $1.646\pm0.006$  
& $1.777\pm0.008$  
& $1.68$ 
\cr
$\pi\Lambda\Sigma^{\ast}$  
& $1.164\pm0.004$  
& $1.178\pm0.006$  
& $1.18$ 
\cr
$\pi\Sigma\Sigma^{\ast}$  
& $-1.164\pm0.004$  
& $-1.059\pm0.007$  
& $-0.68$ 
\cr
$\pi\Xi\Xi^{\ast}$  
& $1.164\pm0.004$  
& $1.111\pm0.007$  
& $\cdots$ 
\cr
$K\Sigma\Delta$  
& $-4.815\pm0.046$  
& $-4.551\pm0.045$  
& $-4.90$ 
\cr
$KN\Sigma^{\ast}$  
& $-3.404\pm0.032$  
& $-3.667\pm0.038$  
& $-2.00$ 
\cr
$K\Xi\Sigma^{\ast}$  
& $3.404\pm0.032$  
& $3.450\pm0.033$  
& $\cdots$ 
\cr
$K\Sigma\Xi^{\ast}$  
& $3.404\pm0.032$  
& $3.167\pm0.032$  
& $\cdots$ 
\cr
$K\Lambda\Xi^{\ast}$  
& $-5.897\pm0.056$  
& $-6.535\pm0.064$  
& $\cdots$ 
\cr
$K\Xi\Omega$  
& $8.339\pm0.079$  
& $8.130\pm0.080$  
& $\cdots$ 
\cr
\hline 
\end{tabular}\label{tab:4}
\end{table}
In Table~\ref{tab:4} we list the results of the pseudovector coupling
constants for the $P_{8}B_{8}B_{10}$ vertices. We find that the present
value of $f_{\pi\Sigma\Sigma^{*}}$ is different from that taken from
the J\"ulich-Bonn potential by almost 50~\%. The value of
$f_{KN\Sigma^{*}}$ differs by approximately 45~\%. However, we want to
emphasize that the present results of the coupling constants reproduce
the experimental data on the decay widths of the decuplet hyperons
very well, which will be discussed now. 

The partial width for the decay from the baryon
decuplet to the octet and pseudoscalar meson $P_8$ is expressed
in terms of the pseudovector coupling constant as follows 
\begin{align}
\Gamma_{B_{10}\to\varphi B_{8}}
\;= & 
\frac{\left|\bm{k}\right|^{3}}{8\pi\,m_{8}^{2}}
\frac{M_{8}}{M_{10}}\;f_{P_8 B_{8}B_{10}}^{2},
\label{eq:decaywidth10-8}
\end{align}
where $|\bm{k}|$ denotes the three momentum of the pseudoscalar meson
in the rest frame of the baryon decuplet. $m_8$ represents the
mass of the pseudoscalar meson involved in the decay process. Summing
all possible transitions with averaging over the initial states, we
can write the decay width for each member of the baryon decuplet as
\begin{align}
\Gamma\left[\Delta\rightarrow\pi N\right] 
& =
\frac{3}{2}\,\Gamma\left[\Delta^{+}\rightarrow\pi^{0}p\right],
\cr
\Gamma\left[\Sigma^{\ast}\rightarrow\pi\Lambda\right] 
& =
\;\;\Gamma\left[\Sigma^{\ast0}\rightarrow\pi^{0}\Lambda\right],
\cr
\Gamma\left[\Sigma^{\ast}\rightarrow\pi\Sigma\right] 
& =
2\,\Gamma\left[\Sigma^{\ast+}\rightarrow\pi^{0}\Sigma^{+}\right],
\cr
\Gamma\left[\Xi^{\ast}\rightarrow\pi\Xi\right] 
& =
3\,\Gamma\left[\Xi^{\ast0}\rightarrow\pi^{0}\Xi^{0}\right].
\label{eq:Widthfactor}
\end{align}
Except for the $\Delta$ decay, the present results are in good
agreement with the experimental data as shown in Table~\ref{tab:5}.
\begin{table}[htp]
\caption{Partial ($\Gamma_{i}$) and full decay widths ($\Gamma$) for
  the decays $B_{10}\to B_{8}+\pi$ in units of MeV. }
\begin{tabular}{ccccc}
\hline 
Decay modes  
& $\Gamma_{i}^{(0)}$  
& $\Gamma_{i}^{\mathrm{(total)}}$  
& $\Gamma$  
& ${\displaystyle \Gamma(\mathrm{Exp.})}$\cite{PDG} 
\cr
\hline 
$\Delta\rightarrow N\pi$  
& $75.98\pm1.01$  
& \multicolumn{2}{c}{$88.58\pm1.31$} 
& $116-120$ 
\cr
\hline 
$\Sigma^{\ast+}\rightarrow$$\Sigma^{0}\pi^{+}$  
& $2.59\pm0.03$  
& $3.22\pm0.06$  
& \multirow{3}{*}{$36.25\pm0.42$}  
& \multirow{3}{*}{$36.0\pm0.7$}
\cr
$\Sigma^{\ast+}\rightarrow$ $\Sigma^{+}\pi^{0}$  
& $3.17\pm0.05$  
& $2.62\pm0.05$  
&  & \cr
$\Sigma^{\ast+}\rightarrow$ $\Lambda\pi^{+}$  
& $29.68\pm0.26$  
& $30.41\pm0.33$  
&  & \cr
\hline 
$\Sigma^{\ast0}\rightarrow$ $\Sigma^{0}\pi^{0}$  
& $0$  & $0$  
& \multirow{4}{*}{$37.21\pm0.69$}  
& \multirow{4}{*}{$36\pm5$}
\cr
$\Sigma^{\ast0}\rightarrow$ $\Sigma^{+}\pi^{-}$  
& $3.61\pm0.11$  
& $2.98\pm0.1$  
&  & 
\cr
$\Sigma^{\ast0}\rightarrow$ $\Sigma^{-}\pi^{+}$  
& $2.78\pm0.1$  
& $2.30\pm0.09$  
&  & 
\cr
$\Sigma^{\ast0}\rightarrow\Lambda\pi^{0}$  
& $31.15\pm0.47$  
& $31.92\pm0.52$  
&  & 
\cr
\hline 
$\Sigma^{\ast-}\rightarrow$ $\Sigma^{-}\pi^{0}$  
& $3.50\pm0.06$  
& $2.89\pm0.06$  
& \multirow{3}{*}{$38.18\pm0.48$}  
& \multirow{3}{*}{$39.4\pm2.1$}
\cr
$\Sigma^{\ast-}\rightarrow$ $\Sigma^{0}\pi^{-}$  
& $3.64\pm0.06$  
& $3.01\pm0.06$  
&  & \cr
$\Sigma^{\ast-}\rightarrow$ $\Lambda\pi^{-}$  
& $31.50\pm0.30$  
& $32.28\pm0.37$  
&  & 
\cr
\hline 
$\Xi^{\ast0}\rightarrow$ $\Xi^{0}\pi^{0}$  
& $4.76\pm0.05$  
& $4.33\pm0.06$  
& \multirow{2}{*}{$11.26\pm0.17$}  
& \multirow{2}{*}{$9.1\pm0.5$}
\cr
$\Xi^{\ast0}\rightarrow$$\Xi^{-}\pi^{+}$  
& $7.61\pm0.08$  
& $6.93\pm0.10$  
&  & 
\cr
\hline 
$\Xi^{\ast-}\rightarrow$ $\Xi^{-}\pi^{0}$  
& $4.76\pm0.05$  
& $4.33\pm0.06$  
& \multirow{2}{*}{$13.01\pm0.21$}  
& \multirow{2}{*}{$9.9_{-1.9}^{+1.7}$}
\cr
$\Xi^{\ast-}\rightarrow$ $\Xi^{0}\pi^{-}$  
& $8.20\pm0.13$  
& $8.68\pm0.16$  
&  & 
\cr
\hline 
\end{tabular}\label{tab:5} 
\end{table}
There exist also experimental data on the ratio of the decay widths
for $\Sigma^{*}\to\Sigma$ and $\Sigma^{*}\to\Lambda$. The present
result is comparable with the data as shown in the following 
\begin{align}
\frac{\Gamma\left[\Sigma^{\ast} \rightarrow \Sigma \right]}{\Gamma
  [\Sigma^{\ast}\rightarrow\Lambda]}  
& =
0.180\pm0.002\;\;\;\left(\mbox{experimental data~\cite{PDG}:
  }0.135\pm0.011\right).
\label{eq:SigmaLBranching}
\end{align}

\begin{table}[htp]
\caption{Pseudovector coupling constants for the $P_{8}B_{10}B_{10}$
  vertices. The second column lists the results for the
  $\mathrm{SU}_f(3)$ symmetric case, whereas the third one does those
  with explicit $\mathrm{SU}_f(3)$ symmetry breaking taken into account.}
\label{tab:6}
\begin{tabular}{ccc}
\hline 
$P_{8}B_{10}B_{10}$  
& $f_{P_{8}B_{10}B_{10}}^{(0)}$  
& $f_{P_{8}B_{0}B_{10}}^{\mathrm{(total)}}$ 
\cr
\hline 
$\pi\Delta\Delta$  
& $0.769\pm0.003$  
& $0.780\pm0.004$ 
\cr
$\pi\Sigma^{\ast}\Sigma^{\ast}$  
& $0.688\pm0.003$  
& $0.703\pm0.004$ 
\cr
$\pi\Xi^{\ast}\Xi^{\ast}$  
& $0.421\pm0.002$  
& $0.469\pm0.002$ 
\cr
$\pi\Omega\Omega$  
& $0$  
& $0$ \cr
$K\Delta\Sigma^{\ast}$  
& $-1.423\pm0.014$  
& $-1.375\pm0.014$ 
\cr
$K\Sigma^{\ast}\Xi^{\ast}$  
& $-2.013\pm0.020$  
& $-2.014\pm0.020$ 
\cr
$K\Xi^{\ast}\Omega$  
& $-2.466\pm0.024$  
& $-2.507\pm0.025$ 
\cr
\hline 
\end{tabular}
\end{table}
In Table~\ref{tab:6}, we list the results on the pseudovector coupling
constants for the $P_8B_{10}B_{10}$ vertices.  The $\pi\Omega\Omega$
coupling constant vanishes, since the isoscalar $\Omega$ baryon can
not be coupled to the pion. Note that as the absolute value of
strangeness increases, the magnitude of the $P_8B_{10}B_{10}$ coupling
constant tends to increase. For example, the magnitude of
$|f_{K\Xi^*\Omega}|$ is approximately three times larger than that of
$f_{\pi\Delta \Delta}$.

\section{Coupling constants for the $\eta$-$B$-$B$,
  $\eta^{\prime}$-$B$-$B$ vetices}

In this Section, we provide the numerical values of the coupling
constants when $\eta$ and $\eta'$ are involved. In order to compute
them, we have to consider the mixing between the octet $\eta_8$ and
the singlet $\eta_0$ coupling constants. Following the mixing scheme
suggested in Ref. \cite{Rijken:2010zzb} given as 
\begin{align}
g_{\eta B_{8}B_{8}}
& =
\mathrm{cos}\theta_{p}\, g_{\eta_8 B_{8}B_{8}}
\;-\;\mathrm{sin}\theta_{p}\, g_{\eta_0 B_{8}B_{8}},
\\
g_{\eta' B_{8}B_{8}} 
& =
\mathrm{sin}\theta_{p}\, g_{\eta_8 B_{8}B_{8}} 
\;+\;\mathrm{cos}\theta_{p}\, g_{\eta_0 B_{8}B_{8}},  
\label{eq:etamixing}
\end{align}
one can easily determine the
coupling constants for the $\eta$ and $\eta'$ coupling
constants. Using the values of $f_{\eta}\,=\,94.0\,\mathrm{MeV}$, 
$f_{\eta^{\prime}}\,=\,89.1\,\mathrm{MeV}$ 
taken from Refs.~\cite{Barnett:1996hr,Behrend:1990sr,Aihara:1990nd} 
and mixing angle $\theta_{p}\,=\, -23.00^{\circ}$ from Ref. \cite{Rijken:2010zzb}, 
we obtain the pseudoscalar coupling constants for the $\eta B_8 B_8$ and
$\eta' B_8 B_8$ coupling constants.

\begin{table}[htp]
\caption{Pseudoscalar strong coupling constants $g_{\eta B_8B_8}$ and
  $g_{\eta' B_8B_8}$ divided by $\sqrt{4\pi}$. The second column lists
  the results for the $\mathrm{SU}_f(3)$ symmetric case, whereas the
  third one corresponds to those from $a_4$, $a_5$, and $a_6$ of the
  collective operator for the axial-vector constants given in 
  Eq.~\eqref{eq:4}. The fourth one represents the corrections from the
  symmetry-breaking parts of the collective wavefunctions in
  Eq.~\eqref{eq:6}. The fifth column presents the total results of the
  coupling constants. The sixth one lists the values of the coupling
  constants taken from the Nijmegen potentials.}  
\label{tab:eta88}
\begin{tabular}{cccccc}
\hline 
$P_{8}B_{8}B_{8}$  
& $g_{P_{8}B_{8}B_{8}}^{\text{(0)}}$  
& $g_{P_{8}B_{8}B_{8}}^{\text{(op)}}$  
& $g_{P_{8}B_{8}B_{8}}^{\text{(wf)}}$  
& $g_{P_{8}B_{8}B_{8}}^{\mathrm{(total)}}$  
& ESC08a \cite{Rijken:2010zzb} 
\cr
\hline 
$\eta NN$  
& $1.583\pm0.126$ 
& $-0.328\pm0.027$ 
& $-0.015\pm0.002$ 
& $1.241\pm0.103$ 
& $1.933$ 
\cr
$\eta^{\prime}NN$  
& $1.241\pm0.103$ 
& $-0.637\pm0.044$ 
& $0.007\pm0.001$ 
& $0.611\pm0.088$ 
& $2.443$ 
\cr
$\eta\Lambda\Lambda$  
& $-1.947\pm0.153$ 
& $1.169\pm0.097$ 
& $-0.053\pm0.005$ 
& $-0.831\pm0.086$ 
& $-1.572$ 
\cr
$\eta^{\prime}\Lambda\Lambda$  
& $3.189\pm0.199$ 
& $2.272\pm0.155$ 
& $0.024\pm0.002$ 
& $5.486\pm0.329$ 
& $4.634$ 
\cr
$\eta\Sigma\Sigma$  
& $3.772\pm0.288$ 
& $-1.026\pm0.086$ 
& $-0.006\pm0.003$ 
& $2.740\pm0.214$ 
& $4.547$ 
\cr
$\eta^{\prime}\Sigma\Sigma$  
& $0.790\pm0.113$ 
& $-2.531\pm0.171$ 
& $0.003\pm0.001$ 
& $-1.738\pm0.173$ 
& $2.168$ 
\cr
$\eta\Xi\Xi$  
& $-3.590\pm0.274$ 
& $1.470\pm0.124$ 
& $-0.042\pm0.004$ 
& $-2.161\pm0.177$ 
& $-2.986$ 
\cr
$\eta^{\prime}\Xi\Xi$  
& $4.346\pm0.266$ 
& $3.747\pm0.253$ 
& $0.019\pm0.001$ 
& $8.111\pm0.484$ 
& $5.981$ 
\cr
\hline 
\end{tabular} 
\end{table}
The corresponding numerical results are listed in
Table~\ref{tab:eta88} and are compared with those from the 
Nijmegen potentials. Since the effects of SU(3) symmetry breaking seem
rather important, we examine the contributions from the SU(3) symmetry
breaking more closely. In the case of exact $\mathrm{SU}_f(3)$
symmetry, the results are very similar to those from the Nijmegen 
potentials. However, when the effects of explicit $\mathrm{SU}_f(3)$
symmetry breaking are taken into account, the values of the $\eta$ and
$\eta'$ coupling constants are in general much changed. As shown in
Table~\ref{tab:eta88}, there are two different contributions of the
$\mathrm{SU}_f(3)$ symmetry breaking: The one arises directly
from the collective operator for the axial-vector constant given in
Eq.~\eqref{eq:4} and the other comes from the wavefunctions mixed with
the states from higher representations as in Eq.~\ref{eq:6}. As
clearly shown in the fourth column of Table~\ref{tab:eta88}, the
wavefunction corrections are negligibly small. However, the linear
$m_s$ corrections from the collective operator, in particular, when it
comes to the $\eta' B_8B_8$ coupling constants, are sizable, even
compared with the contributions of the $\mathrm{SU}_f(3)$ symmetric
terms. 

In order to understand this, we need to examine carefully the
expression for the singlet axial-vector constant $g_A^{(0)}$. As
discussed in detail In Ref.~\cite{Kim:2000vn}, the singlet
axial-vector operator $\hat{g}_A^{(0)}$ is written as 
\begin{align}
\frac{\hat{g}_A^{(0)}}{2}  = a_3 \hat{J}_3 + \sqrt{3} (a_5-a_6) D_{83}^{(8)}, 
\label{eq:singlet}
\end{align}
where the leading-order contribution with $a_1$ vanish. It means that
$a_3$, which is subleading in the $1/N_c$ expansion, plays a leading
role~\cite{Wakamatsu:1993nq, Christov:1993ny}. The parameter $a_3$
in Eq.~\eqref{eq:singlet} comes from the anomalous part of the 
effective chiral action in the $\chi$QSM while in the Skyrme model
it arises from the Wess-Zumino term and vanishes in the version of the
pseudoscalar mesons. Thus, the effects of $\mathrm{SU}_f(3)$ symmetry
breaking are crucial in determining the value of $g_A^{(0)}$
quantitatively. Since $a_5$ and $a_6$ have different signs as shown in
Table~\ref{tab:2}, the $m_s$ correction given in the second term of
Eq.~\eqref{eq:singlet} becomes large. As a result, the effects of
$\mathrm{SU_f(3)}$ symmetry breaking turn out to be sizable, in
particular, in the case of the $\eta' B_8B_8$ coupling constants for
which the singlet contributions are large. Thus, the present results
imply physically that the effects of $\mathrm{SU_f(3)}$ symmetry
breaking are crucial in determining the $\eta$ and $\eta'$ coupling
constants quantitatively. 

\begin{table}[htp]
\caption{Pseudovector coupling constants $f_{\eta B_8B_{10}}$ and
  $f_{\eta' B_8B_{10}}$ divided by $\sqrt{4\pi}$. The second column lists
  the results for the $\mathrm{SU}_f(3)$ symmetric case, whereas the
  third one corresponds to those from $a_4$, $a_5$, and $a_6$ of the
  collective operator for the axial-vector constants given in 
  Eq.~\eqref{eq:4}. The fourth one represents the corrections from the
  symmetry-breaking parts of the collective wavefunctions in
  Eq.~\eqref{eq:6}. The fifth column presents the total results of the
  coupling constants.}
\label{tab:eta810}
\begin{tabular}{ccccc}
\hline 
$P_{8}B_{8}B_{10}$  
& $f_{P_{8}B_{8}B_{10}}^{(0)}$  
& $f_{P_{8}B_{8}B_{10}}^{(\mathrm{op})}$  
& $f_{P_{8}B_{8}B_{10}}^{(\mathrm{wf})}$  
& $f_{P_{8}B_{8}B_{10}}^{\mathrm{(total)}}$ 
\cr
\hline 
$\eta\Sigma\Sigma^{\ast}$  
& $3.21\pm0.25$ 
& $-0.74\pm0.06$ 
& $0.02\pm\pm0.01$ 
& $2.48\pm0.20$
\cr
$\eta^{\prime}\Sigma\Sigma^{\ast}$  
& $3.46\pm0.31$ 
& $-3.00\pm0.20$ 
& $-0.01\pm0.01$ 
& $0.45\pm0.28$
\cr
$\eta\Xi\Xi^{\ast}$  
& $3.21\pm0.25$ 
& $-0.68\pm0.06$ 
& $-0.10\pm0.01$ 
& $2.42\pm0.19$
\cr
$\eta^{\prime}\Xi\Xi^{\ast}$  
& $3.46\pm0.31$ 
& $-3.04\pm0.21$ 
& $0.08\pm0.01$ 
& $0.49\pm0.28$
\cr
\hline 
\end{tabular}
\end{table}

\begin{table}[htp]
\caption{Pseudovector strong coupling constants of the baryon decuplet with
$\eta$ and $\eta^{\prime}$, divided by $\sqrt{4\pi}$. The second column lists
  the results for the $\mathrm{SU}_f(3)$ symmetric case, whereas the
  third one corresponds to those from $a_4$, $a_5$, and $a_6$ of the
  collective operator for the axial-vector constants given in 
  Eq.~\eqref{eq:4}. The fourth one represents the corrections from the
  symmetry-breaking parts of the collective wavefunctions in
  Eq.~\eqref{eq:6}. The fifth column presents the total results of the
  coupling constants.}
\label{tab:eta1010}
\begin{tabular}{ccccc}
\hline 
$P_{8}B_{10}B_{10}$  
& $f_{P_{8}B_{10}B_{10}}^{(0)}$  
& $f_{P_{8}B_{0}B_{10}}^{\mathrm{(op)}}$  
& $f_{P_{8}B_{0}B_{10}}^{\mathrm{(wf)}}$  
& $f_{P_{8}B_{0}B_{10}}^{\mathrm{(total)}}$ 
\cr
\hline 
$\eta\Delta\Delta$  
& $1.77\pm0.15$ 
& $-0.21\pm0.02$ 
& $-0.04\pm0.01$ 
& $1.51\pm0.13$
\cr
$\eta^{\prime}\Delta\Delta$  
& $4.58\pm0.35$ 
& $-0.83\pm0.06$ 
& $0.04\pm0.01$ 
& $3.79\pm0.32$
\cr
$\eta\Sigma^{\ast}\Sigma^{\ast}$  
& $1.16\pm0.10$ 
& $0.02\pm0.01$ 
& $-0.01\pm0.01$ 
& $1.17\pm0.11$
\cr
$\eta^{\prime}\Sigma^{\ast}\Sigma^{\ast}$  
& $5.06\pm0.37$ 
& $-0.01\pm0.01$ 
& $0.01\pm0.01$ 
& $5.05\pm0.37$
\cr
$\eta\Xi^{\ast}\Xi^{\ast}$  
& $0.56\pm0.07$ 
& $0.21\pm0.02$ 
& $-0.03\pm0.01$ 
& $0.75\pm0.08$
\cr
$\eta^{\prime}\Xi^{\ast}\Xi^{\ast}$  
& $5.53\pm0.39$ 
& $0.83\pm0.06$ 
& $0.02\pm0.01$ 
& $6.39\pm0.43$
\cr
$\eta\Omega\Omega$  
& $-0.04\pm0.05$ 
& $-0.22\pm0.02$ 
& $-0.01\pm0.01$ 
& $-0.27\pm0.06$
\cr
$\eta^{\prime}\Omega\Omega$  
& $6.00\pm0.41$ 
& $-0.83\pm0.06$ 
& $0.01\pm0.01$ 
& $5.18\pm0.38$
\cr
\hline 
\end{tabular}
\end{table}
In Table~\ref{tab:eta810}, we list the results of the $\eta B_8 B_{10}$ and
$\eta' B_8 B_{10}$ coupling constants. In general, the effects of explicit
$\mathrm{SU}_f(3)$ symmetry breaking reduce the magnitudes of these
coupling constants noticeably. 
Table~\ref{tab:eta1010} lists the results of the $\eta$ and $\eta'$
coupling constants for the baryon decuplet. Interestingly, 
the effects of explicit $\mathrm{SU}_f(3)$ symmetry breaking are
marginal except for the $\eta \Omega \Omega$ coupling constant, since
the matrix elements of $D_{83}^{(8)}$ are small for the baryon
decuplet. Note that the $\eta' \Sigma^* \Sigma^*$ does not acquire any 
contribution from explicit $\mathrm{SU}_f(3)$ symmetry breaking. 
In general, the values of the $\eta'$ coupling constants are much
larger than those of the $\eta$ ones. 

\section{Summary and conclusion}

In the present work, we have investigated the strong coupling constants
for the meson-baryon-baryon vertices within the general framework
of the chiral soliton model, taking into account the effects of flavor
SU(3) symmetry breaking to linear order. All the relevant dynamical
parameters were fixed by using the experimental data on the hyperon
semileptonic decays and the singlet axial-vector constant. We were
able to determine the strong coupling constants for the baryon octet
and pseudoscalar meson octet vertices, those for the transition from
the baryon decuplet to the baryon and pseudoscalar meson octets, and
those for the baryon decuplet and pseudoscalar meson octet vertices.
Except for the $\pi\Xi\Xi$, $\pi\Sigma\Sigma^{*}$ and $KN\Sigma^{*}$
vertices, the present results were in good agreement with those determined
from the Nijmegen and J\"ulich-Bonn potentials. We also computed the
decay widths of the baryon decuplet to the baryon octet and the pion.
Apart from the $\Delta$ decays, the results are in good agreement
with the experimental data. We also presented the strong coupling
constants for the $\eta$ and $\eta'$ mesons. The effects of
$\mathrm{SU_f(3)}$ symmetry breaking are in general quite sizable on
the $\eta'$ coupling constants. This can be understood that the
leading contribution to the singlet axial-vector constant vanishes in
the $1/N_c$ expansion within the present framework and the
subleading-order terms play a leading role. Thus, the corrections of
the strange current quark mass become relatively more important in the
case of the $\eta'$ coupling constants. 

The strong coupling constants for the vector mesons and the baryon
octet and decuplet can be examined within the same framework. Since
the vector mesons have spin 1, the structure of the coupling constants
is more involved. The related work is under investigation.

\section*{Acknowledgments}
H.-Ch.K is grateful to M. V. Polyakov for the discussion and
hospitality during his visit to the Institute f\"ur Theoretical
Physics II, Ruhr-Universit\"at Bochum, where part of the work was
done. The present work was supported by Basic Science Research Program
through the National Research Foundation of Korea funded by the
Ministry of Education, Science and Technology (Grant
No. NRF-2016R1C1B1012429 (Gh.-S. Y.) and 2018R1A2B2001752(H.-Ch.K.)). 

\end{document}